\documentclass[conference]{IEEEtran}
\IEEEoverridecommandlockouts
\usepackage{cite}
\usepackage{amsmath,amssymb,amsfonts}
\usepackage{algorithmic}
\usepackage{algorithm}
\usepackage{graphicx}
\usepackage{textcomp}
\usepackage{xcolor}
\usepackage{multirow}
\usepackage{siunitx}
\usepackage{subcaption}
\usepackage{booktabs}

\def\BibTeX{{\rm B\kern-.05em{\sc i\kern-.025em b}\kern-.08em
    T\kern-.1667em\lower.7ex\hbox{E}\kern-.125emX}}
\begin{document}

\title{Autonomous Decision Making \\for Air Taxi Networks\\

\author{\IEEEauthorblockN{Alex Vesel}
avesel@stanford.edu}
}

\maketitle

\begin{abstract}
Future urban air mobility systems are expected to be operated by rideshare companies as fleets, which will require fully autonomous air traffic control systems and an order of magnitude increase in airspace capacity. Such a system must not only be safe, but also highly responsive to customer demand. This paper proposes the air traffic network problem (ATNP), which models the optimization problem of future cooperative air taxi networks. We propose a three-phase decision making model that efficiently assigns vehicles to passengers, determines flight levels to reduce collision risk, and resolves aircraft conflicts by selectively applying Monte Carlo tree search. We develop a simulator for the ATNP and show that our approach has increased safety and reduced passenger waiting time compared to greedy and first-dispatch protocols over potential vertiport layouts across the Bay Area and New York City. 
\end{abstract}

\section{Introduction}
As the world sees an increasing trend towards urbanization, drivers in cities are spending more time in traffic, with the average American driver losing 51 hours and \$869 annually due to road congestion \cite{owid-urbanization}, \cite{Inrix_2023}. The cost is most acute in urban areas, such as Chicago, New York, and the Bay Area, that experience high levels of commuter traffic. One potential solution to this problem is Urban Air Mobility (UAM) enabled by the development of electric vertical take-off and landing (eVTOL) aircraft. Based out of hubs called vertiports, these aircraft will travel 10 to 100 miles per flight and collectively service hundreds to thousands of people per hour \cite{mckinsey}.

To characterize the development and potential future direction of UAM systems, NASA developed a framework called the the UAM Maturity Level (UML) scale \cite{Goodrich2021}. These levels 1-6 characterize increasing availability, complexity, and autonomy of potential UAM systems. For example, UML 1 describes eVTOLs in the testing phase while UML 6 describes fully autonomous and ubiquitous air travel characterized by ad-hoc landing areas such as driveways and neighborhood streets \cite{Goodrich2021}. This work considers a UML 5 system, which features a network of hundreds of unmanned aerial vehicles (UAVs) operating in a given metro area.

However, the introduction of hundreds to thousands of eVTOLs operating in close proximity will require the development of significant new air traffic management (ATM) and autonomous flight technology \cite{Xiang2023}. While efforts such as the Federal Aviation Administration's (FAA) Next Generation Air Transportation System (NextGen) aim to increase air traffic capacity through new aircraft surveillance technology and increased reliance on ATM algorithms, the agency's current plan will require eVTOLs to operate under visual flight rules with human pilots, which will not scale to UML 5 systems \cite{nextgen}, \cite{nextgen_evtol}. 

Furthermore, due to the high expected cost of eVTOLs, it is anticipated that the majority of UAM aircraft will be operated by professional operators as air taxis, shuttling passengers between vertiports \cite{Vascik2020SystemsAO}, \cite{Rajendran2020}. These UAV ridesharing services will face similar dynamic scheduling problems as automotive ridesharing services such as Uber, with fleet capacity utilization and trip throughput being key performance metrics \cite{ubermatching}. However, in contrast to the automotive ridesharing scenario where the matching of drivers to passengers and route planning of vehicles can be treated as independent problems, collision avoidance constraints in the UAM ridesharing scenario require matching and path planning to be considered jointly. For example, consider a myopic matching between a large number of air taxis and passengers waiting at vertiports. As the aircraft proceed towards their target destinations, the high density of airspace will require many collision avoidance deviations from the optimal individual flight paths. As a result of these deviations, the optimal myopic matching will change, resulting in network inefficiencies. A system that accounts for both safety and operational efficiency during flight path planning will be critical in enabling a UML 5 future where hundreds of eVTOLs must safely navigate and serve customers as part of a dense and dynamic air taxi network.

This research introduces a new decision making problem called the \textit{air taxi network problem} (ATNP), which describes an optimization problem of how a cooperative fleet of UAVs should generate flight trajectories to serve customers in a network of urban vertiports while maintaining safety. At a high level, the optimization objective will be to maximize the trip throughput of the system, subject to avoiding collision events. This emulates the problem that future air taxis services will have when managing their fleet. We then formulate the ATNP as a multiagent Markov decision process (ATNP-MMDP) and provide a three-phase algorithm to practically solve the ATNP-MMDP. The proposed algorithm is evaluated in a simulation environment that implements the logic of the ATNP in potential vertiport networks across the San Francisco Bay Area and New York City. Code for the simulator and methods can be found on GitHub \footnote{https://github.com/alex-vesel/uam-planning}\label{github}.

The primary contributions of this work are:
\begin{enumerate}
    \item A formulation of the air taxi network problem, which models the optimization problem of future air taxi services
    \item A solution to the ATNP that decomposes the problem into three sub-problems: agent-passenger assignment, flight level selection, and flight trajectory planning.
    \item A simulator of the ATNP grounded in potential vertiport layouts across the Bay Area and New York City.
\end{enumerate}

\section{Background}
\begin{figure*}
    \centering
    \includegraphics[width=0.85\linewidth]{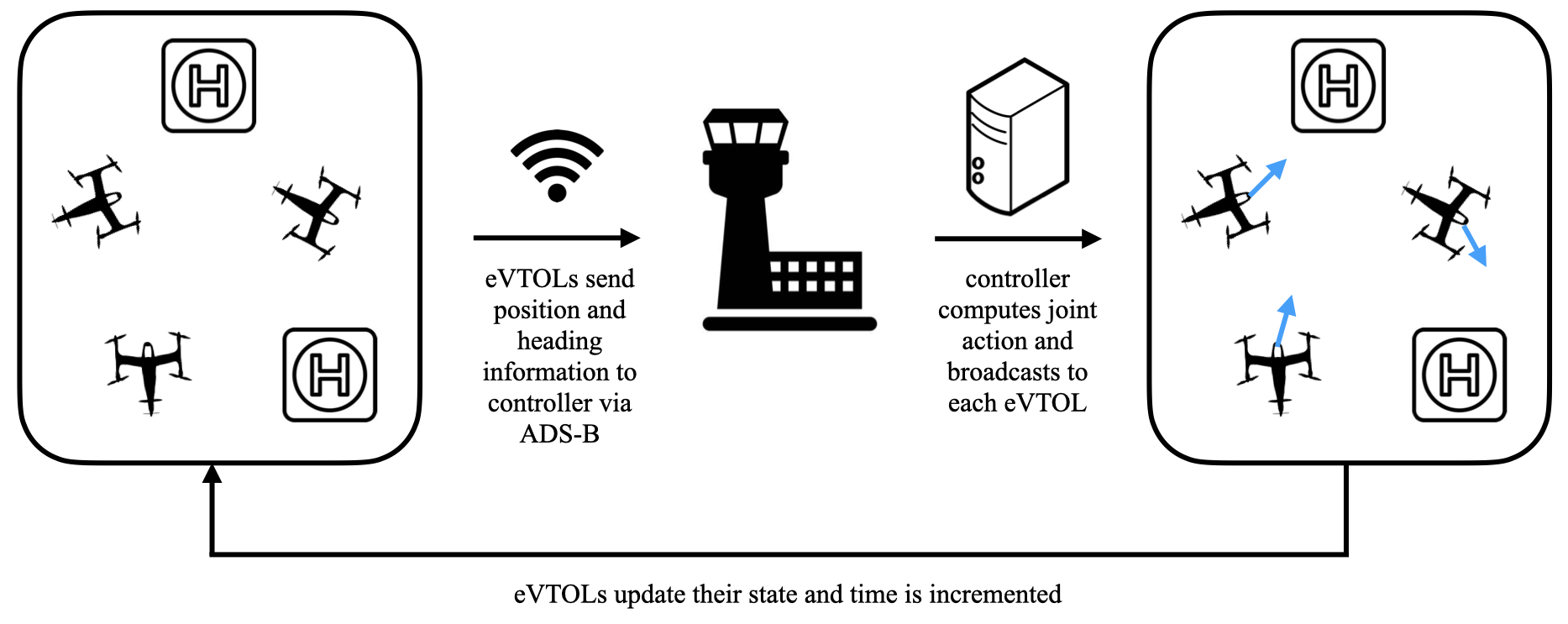}
    \caption{A high level system architecture diagram. At each timestep, the fleet of eVTOLs broadcast their position and heading information to the controller via ADS-B. The controller computes a joint action and broadcasts back to each eVTOL. Each eVTOL takes the joint action over a timestep, resulting in a new map. This process repeats until all passengers are delivered.}
    \label{fig:system-architecture}
\end{figure*}
Related to the proposed ATNP are the problems of vehicle routing, multiagent pathfinding, aircraft collision avoidance, and rideshare matching. Selected prior work from these fields are described in this section.

\subsection{Vehicle Routing Problem}
The vehicle routing problem (VRP) is the NP-Hard combinatorial optimization problem of determining the routes of a fleet of vehicles to visit a set of locations while respecting constraints such as collision avoidance \cite{vrp}. Related is the multi-agent pathfinding (MAPF) problem, which is the computation of optimal paths for a fleet of agents to their assigned targets with respect to some cost function.

VRPs/MAPF often arise in UAV management \cite{Wang2019}. \cite{9197313} considers the problem of a cooperative fleet of delivery drones that can both fly and hitch rides on public transportation to conserve energy. They develop a two-level solution: the first phase assigns determines package starting locations and assigns delivery sequences to drones, and the second uses MAPF techniques to determine routes each drone should take. A similar two-level architecture is used in \cite{Han2023}, which considers a package delivery problem with dynamic delivery conditions. \cite{Yang2020ScalableMC} uses a decentralized partially observable Markov decision processes (Dec-POMDP) to model state transition uncertainty in a network of UAVs for conflict-free trajectory planning. To improve computational tractability, they sectorize the airspace and only plan for collisions within a sector. 

The ATNP can be considered a dynamic vehicle routing problem in that passenger arrivals and destinations are not known in advance, meaning that future demand must be considered at the current timestep with replanning required as new passengers enter the environment \cite{Kucharska2019}, \cite{OjedaRios2021}. Likewise, the ATNP allows for aircraft to operate in free flight with continuous state spaces, precluding the direct application of graph-based MAPF techniques as in \cite{9197313}, although approximation or postprocessing techniques could be employed to enable such an approach \cite{mapf-kc}.

\subsection{Aircraft Collision Avoidance}
Aircraft collision avoidance is a short-term path planning problem that seeks to reconcile potential collisions between $>1$ aircraft heading towards preassigned destinations.

Many prior methods for aircraft collision avoidance assume pairwise aircraft encounters with uncertain state information, with aircraft relying on onboard electro-optical/infrared and radar sensors to detect the intruder aircraft. These approaches often use partially observable Markov decision processes (POMDPs) to model the trade-off between accurately localizing the intruder aircraft and avoiding a collision \cite{Bai2011}, \cite{Wolf2011}, \cite{Mueller2016}. The ACAS X collision avoidance algorithm uses such an approach and is expected to become required for commercial aircraft \cite{acasx2}, \cite{acasx}.

Other work considers collision avoidance between multiple aircraft. \cite{Ong2017} formulates collision avoidance as a multiagent Markov decision process (MMDP) by computing an offline policy that fuses individual agent solutions. \cite{Zhao2020OnlineMC} considers a future risk map of potential collisions and minimizes the risk of collisions using Dijkstra's algorithm. Another line of work uses reinforcement learning to make use of the Centralized
Training with Decentralized Execution paradigm, which learns a policy offline using privileged simulation information to reduces online compute costs for decentralized path finding systems \cite{huang2022multiuav} \cite{Chen2016DecentralizedNM}.

These prior works could be characterized as \textit{decentralized} in that each agent only observes its own local environment and computes a policy without direct coordination with other agents. While modeling the problem in a decentralized manner with uncertainty in the position of the intruder aircraft has been necessary in the past, the widespread adoption of the Automatic Dependent Surveillance-Broadcast (ADS-B) system will allow aircraft to share accurate position and velocity information with the controller and other aircraft at a rate of 1 Hz \cite{equipadsb}, \cite{adsbspec}. In this work, we assume that each eVTOL's position and heading are fully known to the controller through ADS-B. A more complete description of eVTOL communications can be found in \cite{Zaid2023}.

\subsection{Rideshare Matching}
Due to the rise in popularity of automotive rideshare services such as Uber and Lyft, there is extensive literature on the problem of \textit{matching}, which is the process of dispatching available drivers to pick up riders \cite{ubermatching}. At any given time, drivers are part of one of two sets: \textit{open}, meaning they are not currently assigned to any passenger, or \textit{on-trip} which means they are either actively en route to pick up a passenger or are driving a passenger to their destination. One of the simplest approaches to matching is the first-dispatch protocol, where a passenger is immediately assigned to the nearest open driver. However, this approach is suboptimal in that once a driver is committed to picking up a passenger they are not considered for potentially more efficient matchings to new passengers in future timesteps. A technique known as \textit{batching} is often used to mitigate this problem, whereby new requests are held in a pool for some waiting period before being assigned to drivers \cite{Zhang2017}, \cite{ashlagi2018maximum}, \cite{ubermatching}. Solutions to the batching technique can be found efficiently using the Hungarian algorithm \cite{Kuhn1955}. These approaches can be considered myopic in that they do not account for future demand. Other works found that methods that plan for future demand can significantly outperform first-dispatch variants when there are large regional imbalances in supply and demand \cite{zkan2020}.

The requirements of the proposed ARNP differ from the automotive rideshare matching problem in several ways. Firstly, in many automotive rideshare matching problems once a driver is committed to picking up a passenger they remain committed even if preferred matchings arise in the future. This is to better align with the expectations of human drivers who act as independent contractors and passengers who often become informed in advance who their driver will be. However, in the ARNP, air taxis are fully autonomous and operated by the rideshare company, meaning agents can readily be reassigned if better matchings are found in the future. Likewise, the matching system must consider the trajectories agents will follow to reduce the risk of collisions.

\section{Problem Statement}
 This section introduces the proposed Air Taxi Network Problem. We consider an environment of $n$ agents representing eVTOLs and $m$ vertiports which generate $p$ passengers over some horizon. We refer to the configuration of vertiports relative to a common 2D coordinate system collectively as a \textit{map}. Figure \ref{fig:system-architecture} shows the high-level communication architecture of the ATNP. At each timestep, with granularity $\Delta t$, a \textit{centralized controller} selects actions for each agent to maximize the rate at which passengers are delivered to their destinations, subject to a set of feasibility and safety constraints. The environment is described formally as follows.
 
\subsection{Agents}
Each agent $i \in \{1, 2, \dots, n\}$
configuration at time $t$, $g_i^t$, can be specified by a tuple $g_i^t = ( x_i^t, y_i^t, \theta_i^t, k_i^t, p_{i}^t, f^t)$. The part $(x_i^t, y_i^t, \theta_i^t) \in SE(2)$ describes the 2D $x$ and $y$ position of the agent on the map, along with its heading angle $\theta$ with respect to positive $x$ direction. An integer $k_i^t \in \{0, 1, 2, \dots, m\}$ indicating which vertiport the agent is grounded at, where 0 represents the agent is flying. An integer $p_{i}^t \in \{0, 1, 2, \dots, p\}$ indicates which passenger the aircraft is currently carrying, where 0 represents the agent not containing any passenger. $f^t \in \{1, 2, \dots, F\}$ indicates the flight level of the eVTOL from a discrete set of $F$ available flight levels. For the purposes of modeling, we assume each agent travels at the same altitude and velocity $v$, and as such do not include these components in the agent configuration. However, we note that the problem description is readily extendable to the full 3D case. 

At each timestep $t$, each agent $i$ takes an action $a_i^t \in A_i$, with different actions available depending on if the agent is flying or grounded. If the agent is flying, the action is $a_i^t = (\omega_i^t, l)$, where $\omega$ is the desired angular velocity and $l$ is a boolean value indicating if the agent will land at the nearest vertiport. We impose a maximum angular velocity such that $|\omega_i^t| \leq \omega_{\text{max}}$ The agent follows the following kinematic rules when flying:
\[
\theta_i^{t + \Delta t} = \theta_i^t + \omega_i^t \Delta t
\]
\[
x_i^{t + \Delta t} = x_i^{t} + v \cos({\theta_i^{t + \Delta t}}) \Delta t
\]
\[
y_i^{t + \Delta t} = y_i^{t} + v \sin({\theta_i^{t + \Delta t}}) \Delta t
\]
An agent has the ability to land at a vertiport when its distance to the vertiport is less than some threshold $d_{\text{land}}$. We assume the eVTOLs follow some landing protocol, which is beyond the scope of this work but is considered in \cite{8569645} and \cite{Song2022}. When landed, the agent's action is $a_i^t = 
 (\theta_{\text{takeoff}}, s)$, where $\theta_{\text{takeoff}}$ is the desired takeoff angle of the agent and $s$ is a boolean indicating if the agent will remain grounded at the vertiport for the duration of the current timestep. As a simplifying assumption (and motivated for passenger comfort), the agent must remain at the same flight level for the duration of a given flight, meaning the flight level may only be changed once an agent lands and takes off.

 A pair of agents is considered in conflict in two scenarios, termed oss of separation (LOS) and near midair collision (NMAC). LOS is when two aircraft are within 0.5 nautical miles (0.926 km) of each other and NMAC is when two aircraft are within 500 feet (0.15 km) of each other.

\subsection{Vertiports}
Each vertiport $k \in \{1, 2, \dots, m\}$ is parameterized by a tuple $\{ x_k, y_k, \lambda_k \}$, where $x_k$ and $y_k$ indicate the vertiport's position on the map and $\lambda_k$ parameterizes a Poisson distribution $\text{Pois}(\lambda_k)$ describing the hourly arrival rate of passengers. At each  time $t$, each vertiport $k$ contains a set of waiting passengers $P_k^t$. This framework is readily extendable to more complex vertiport models including finite landing pads and modulating passenger arrival rates, as described in \cite{Li2020}.

\subsection{Passengers}
Passengers are randomly generated according to the arrival rates of each vertiport. Each passenger $j$ is described by $(o_j, d_j)$, where $o_j$ is the vertiport where $j$ was generated and $d_j$ is the destination vertiport, which is chosen uniformly at random from the set of all vertiports. A passenger will remain at the origin vertiport until it is picked up by an agent. The passenger's journey is considered complete when the agent lands at the destination vertiport. 

\subsection{ATNP as a Multiagent Markov Decision Process}
A multiagent Markov decison process (MMDP) is a tuple $(\mathcal{N}, \mathcal{S}, \mathcal{A}, \mathcal{T}, \mathcal{R})$ where $\mathcal{N}$ is a set of $n$ agents, $\mathcal{S}$ is the state space, $\mathcal{A} = A_1 \times A_2 \times \dots \times A_n$ is the joint action space, $\mathcal{T}(s, a, s') : S \times A \times S \rightarrow [0, 1]$ is a transition function from current states and joint actions to next states, and $\mathcal{R} : S \times A \rightarrow \mathbb{R}$ is the reward function \cite{mmdp}. One feature of a MMDP is that the reward function is shared between agents. 

We note that the ATNP can be approximated as a multiagent Markov decision process, which we term the ATNP-MMDP. The state space is the Cartesian product of all agent and vertiport configurations $g_1 \times g_2 \times \dots \times g_n \times P_1 \times P_2 \times \dots \times P_k$ and the action space is joint action space $A_1 \times A_2 \times \dots \times A_n$. The transition function is deterministic over the agent subspace of the state space, as defined by the dynamics described in the previous subsection, and stochastic over the waiting passenger lists, where the transitions are defined according to the joint distribution over each vertiport's Poisson arrival time distribution. Now consider two functions over the natural numbers: $d : \mathcal{S} \rightarrow \mathbb{N}$ indicating the number of newly delivered passengers at the input state and $c : \mathcal{S} \rightarrow \mathbb{N}$ indicating the number of collision violations in the input state. The reward function of the MMDP is defined as $R(s, a) = d(s) - \gamma c(s)$, where $\gamma \ge 0$ is a term that converts the collision avoidance constraint in the ATNP to a penalty.

\section{Method}
While formulating the ATNP as a MMDP provides a useful framework, solving MMDPs exactly is often impractical due to the joint action space growing exponentially with the number of agents. Previous planning methods often factor the action and/or reward functions to find an approximate solution \cite{Ong2017}, \cite{Yang2020ScalableMC}. These methods often use online MDP planning algorithms such as Monte Carlo tree search (MCTS) to dynamically plan flight trajectories. However, these prior methods are not well suited for solving the ATNP-MMDP for a few reasons. Firstly, factoring the full space of aircraft encounters into pairwise encounters may be suited for low-density airspace where a pair of agents deviating to avoid collision will not cause further collisions with other aircraft, but the full space of possible future collisions must be considered in the high-density ATNP. Secondly, most prior works assume that an agent's destination is provided a priori, whereas in the ATNP the controller has autonomy to decide where the fleet of agents should be allocated to maximize passenger throughput. Finally, directly applying MCTS methods to the ATNP will not yield useful trajectories due to the sparsity of the reward function: the system must consider plans on the order of 10s of minutes to adequately account for passenger trip durations and the stochastic nature of passenger arrivals. If $\tau_{max}$ specifies the planning horizon, then the system must naively consider at least $|A| ^{n^{\frac{\tau_{max}}{\Delta t}}}$ possible trajectories, not even accounting for future passenger arrivals. This is clearly intractable even for a small discrete action space, a few agents, and a modest planning horizon as only a small subset of possible plans will successfully deliver passengers. 

To address this problem, the proposed solution method decomposes the ATNP into three sub-problems: agent-passenger assignment, flight level selection, and flight trajectory planning. At each timestep, the agent-passenger assignment phase assigns a target passenger for each free agent to attend to. After each agent has been assigned a passenger, the target vertiports for each agent are assigned accordingly given each passenger's origin. Flight level selection determines the flight level for a grounded agent before takeoff. Finally, the flight trajectory planning phase attempts to navigate each agent to its destination while avoiding collisions. Further algorithm details are described in the following subsections.

% While the bipartite matching problem has worst case complexity $O(n!)$, efficient solutions can be found using the Hungarian algorithm. 

\subsection{Agent-Passenger Assignment}
Algorithm \ref{alg:matching} shows the proposed matching procedure, which can be broken into two sub-phases. In the first sub-phase, candidate matchings are found using Murty's algorithm, which provides an efficient solution to the k-best assignments problem. Then the final matching is selected from the candidates by minimizing the L1 norm between the location distribution of agents assuming they deliver all the passengers in the candidate matching and the optimal distribution of agents given the arrival rates of each vertiport.

\subsubsection{Candidate Matching Generation}
A cost matrix $C$ is computed between agents and current waiting passengers such that $C_{ij}$ contains the distance between agent $i$ and passenger $j$'s origin vertiport. We consider all agents in the matching process, even those that are currently on-trip with a passenger (whose entry $C_{ij}$ is the sum of the distance to their current passenger's destination and from the destination to $j$'s origin). The reasoning is that if an agent is about to deliver a passenger to a vertiport with a waiting passenger, there is likely no need to assign a more distant agent to fly there. The top $k$ candidate matchings that minimize the assignment over the cost matrix can be found efficiently using Murty's algorithm \cite{532080bb-34c4-3c21-a5ce-6f11e19925ad}. Empirically, many of the top matchings result in the same agent-vertiport assignments but with different agent-passenger assignments. 

\subsubsection{Final Matching Selection}
We use a simple heuristic to determine which candidate matching to select as the final matching. We introduce a desired distribution of agents to vertiports, which we set proportional to the arrival rates of each vertiport.  For each candidate matching $m$, we determine the distribution of the number of agents at each vertiport if each agent delivers their assigned passenger under matching $m$. We select as the final matching the matching that minimizes the L1 norm between the desired distribution and candidate matching distribution. The intuition behind this heuristic is that we want to favor matchings that best account for future demand.

\begin{algorithm}
\caption{\textsc{Matching}}\label{alg:matching}
\begin{flushleft}
\textbf{Input: } desired agent nearest vertiport distribution $\psi^*$, number of candidate matchings $k$\\
\textbf{Output: } agent-passenger assignment
\end{flushleft}
\begin{algorithmic}[1]
\STATE $C \gets \textsc{GetCosts}$()
\STATE $\texttt{best\_score} \gets \inf$
\FOR{$\text{i} \in \{1, 2, \dots, k\}$}
    \STATE $\texttt{asgmt}_\texttt{i} \gets $\textsc{Murty}(C, i) \hfill\COMMENT{i'th best matching}
    \STATE $\psi_i \gets \textsc{FutureDistribution}(\texttt{asgmt}_\texttt{i})$
    \STATE $\texttt{score} \gets || \psi^* - \psi_i ||_1$
    \IF{$\texttt{score} < \texttt{best\_score}$}
        \STATE $\texttt{best\_asgmt} \gets \texttt{asgmt}_\texttt{i}$
        \STATE $\texttt{best\_score} \gets \texttt{score}$
    \ENDIF
\ENDFOR

\RETURN \texttt{best\_asgmt}

\end{algorithmic}
\end{algorithm}

\begin{algorithm}
\caption{\textsc{GetCosts}}\label{alg:get_costs}
\begin{flushleft}
\textbf{Input: } $n$ agent tuples, $p$ waiting passenger tuples\\
\textbf{Output: } $C \in \mathbb{R}^{n \times p}$
\end{flushleft}
\begin{algorithmic}[1]
\FOR{$\text{i} \in \{1, 2, \dots, n\}$}
    \FOR{$\text{j} \in \{1, 2, \dots, p\}$}
        \STATE \{ $i$ has a passenger \}
        \IF{$p_i != 0$}
    \STATE $C_{ij} \gets \sqrt{(x_i - x_{d_{p_i}}) + (y_i - y_{d_{p_i}})} \newline + \sqrt{(x_{d_{p_i}} - x_{o_j}) + (y_{d_{p_i}} - y_{o_j})}$
        \ELSE
            \STATE $C_{ij} \gets \sqrt{(x_i - x_{o_j}) + (y_i - y_{o_j})}$
        \ENDIF
    \ENDFOR
\ENDFOR

\RETURN C
\end{algorithmic}
\end{algorithm}

\subsection{Flight Level Selection}
\begin{figure*}[ht!]
    \centering
    \begin{subfigure}[t]{0.33\textwidth}
        \centering
        \includegraphics[height=1.6in]{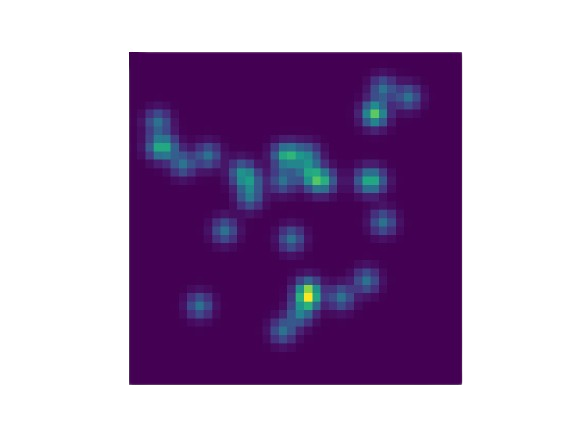}
        \caption{1 step}
    \end{subfigure}%
    \begin{subfigure}[t]{0.33\textwidth}
        \centering
        \includegraphics[height=1.6in]{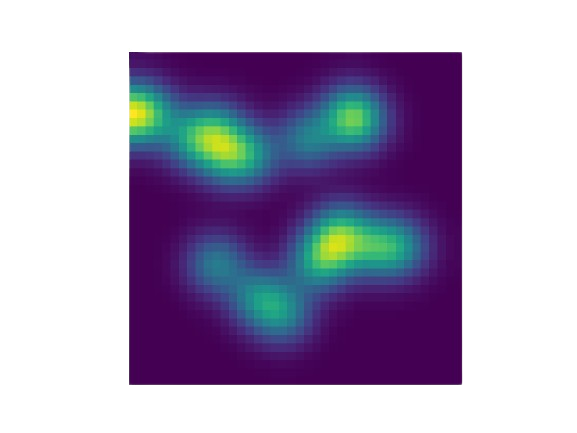}
        \caption{10 steps}
    \end{subfigure}
    \begin{subfigure}[t]{0.33\textwidth}
        \centering
        \includegraphics[height=1.6in]{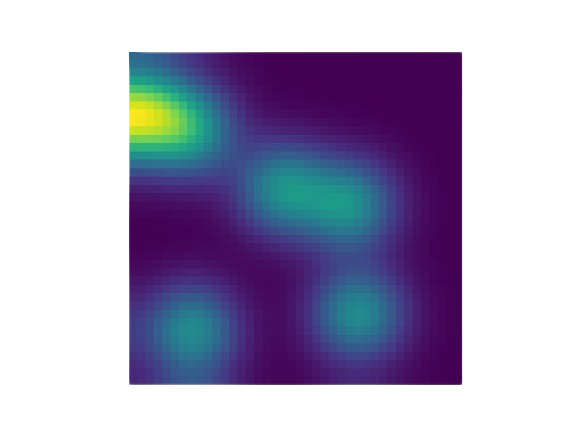}
        \caption{20 steps}
    \end{subfigure}
    \caption{An example of aircraft density grids used in flight level selection for 1, 10 and 20 steps into the future. Note that as the number of steps increase, the densities becomes more diffuse to account for uncertainty in the future locations of each aircraft.}
    \label{fig:heatmaps}
\end{figure*}

In the ATNP, an aircraft maintains a constant flight level for the duration of a flight. Flight level selection chooses a flight level for an agent before takeoff to minimize the risk of potential collisions. We consider a lookahead of $\phi$ steps where each agent follows a greedy policy towards their destination (see Section \ref{greedy_policy}). For each flight level and each lookahead step, a discretized 2D density grid is maintained with the locations of each non-grounded agent. The future location of each agent is modelled as a Guassian distribution, with variance increasing with respect to the lookahead step number to account for uncertainty in the future positions due to potential deviations. Figure \ref{fig:heatmaps} shows an example of these density grids for 1, 10, and 20 lookahead steps. The total risk for agent $i$ being assigned to a flight level is defined as the sum of density along $i$'s future greedy trajectory. We iterate over each grounded agent and select the flight level that minimizes the total risk score. Finally, $i$'s trajectory is added to the selected flight level's density grid before the process repeats for agent $i+1$.

\subsection{Flight Trajectory Planning}
The trajectory planner maps the current agent states and assigned target vertiports/passengers to actions for each agent. We consider two types of trajectory planning: greedy planning, which takes the action that minimizes the flight time to the destination regardless of potential collisions, and MCTS planning which refines the greedy plan to avoid collisions.

\subsubsection{Greedy Planning}\label{greedy_policy} The greedy plan does not take into account potential collisions, meaning each agent's action $a_{i, \text{greedy}}^t$ can be computed independently from other agents. If the agent is in flight, the heading of the target vertiport with respect to the agents current heading $\theta_{\text{target}}$ is calculated. The agent then selects the angular velocity $\omega_{\text{greedy}}$ that minimizes this desired heading angle, within the constraint of $|\omega_{\text{greedy}}| \leq \omega_{\text{max}}$. If the agent is within range of the target vertiport, their action will be to land, and if the agent is grounded and the target vertiport their action will be to stay grounded, potentially resulting in a new passenger being loaded. If the agent is ready to takeoff from a vertiport, it selects the takeoff angle as $\theta_{\text{takeoff}} = \theta_{\text{target}}$. The joint action is simply the concatenation of the greedy actions for each agent $a_{\text{greedy}}^t = [a_{1, \text{greedy}}^t, a_{2, \text{greedy}}^t, \dots, a_{n, \text{greedy}}^t]$

\subsubsection{MCTS Planning}
The MCTS planning used in this work employs a number of heuristics to find locally optimal solutions, using techniques also found in \cite{Ong2017} and \cite{Yang2020ScalableMC}. For MCTS, we fix a discrete set of actions to search over. The joint action space has exponential complexity with respect to the number of agents $n$, we use alternating maximization where for agent $i$ we fix the actions for all other agents and find the action $a_{i, \text{MCTS}}^t$ that maximizes reward for agent $i$. We then insert the updated action into the joint action vector and repeat the process for agent $i+1$, $i+2$, etc. Each agent attempts to maximize an individual reward where at each search node the reward is inversely proportional to the distance of agent $i$ to its assigned target if the agent is flying, some large negative number $r_{\text{LOS}}$ if the agent has lost separation with another agent, and some large positive number $r_{\text{land}}$ if the agent lands at its target destination. The joint action considered at the start of the search is the greedy policy $a_{\text{greedy}}^t$. The MCTS algorithm used in this work is the UCT algorithm \cite{uct}.

To improve computation speed, MCTS is only run on agents that will be in LOS in a greedy $\Delta t_{\text{lookahead}}$ rollout of the current state, given the assigned targets. That is, if the agents' targets remain the same over the next $\Delta t_{\text{lookahead}}$ seconds and they follow the greedy policy during that duration, MCTS is only run on the agents that have LOS during this rollout. This drastically reduces computation time as in any given state the number of agents that might have LOS over $\Delta t_{\text{lookahead}}$ is much lower than $n$. This also follows the approach used in many aircraft collision avoidance works, where aircraft typically follow the optimal path to their destination and are only issued diversions during an encounter with another aircraft \cite{Bai2011}, \cite{Wolf2011}. 

The final heuristic modifies the MCTS algorithm itself. During an iteration of alternating maximization for agent $i$, the MCTS only branches on actions for agent $i$, requiring a method of generating a joint action at each tree search node. \cite{Yang2020ScalableMC} considers fully random joint actions of other agents at each MCTS node. However, fully random actions leads to very conservative policies, meaning that agents are assigned wider deviations than necessary. Given that our policy will assign greedy actions when there is no potential collision, we can select the greedy joint action at each MCTS node. Empirically this modification results in a more efficient policy.

\section{Simulation Framework}
We develop a simulator that implements the ATNP in Python, introduce a set of evaluation metrics, and run experiments over varying numbers of eVTOLs and vertiports.

\subsection{Simulator}
\begin{figure}[t]
\centering
\begin{subfigure}[b]{1\linewidth}
    \centering
   \includegraphics[width=0.8\linewidth]{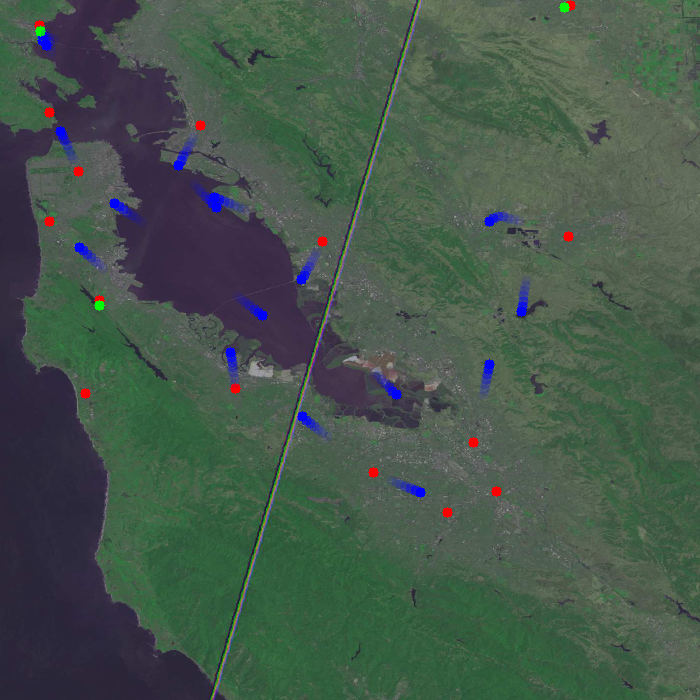}
   \caption{Bay Area Map}
   \label{fig:frame_sf} 
\end{subfigure}

\begin{subfigure}[b]{1\linewidth}
    \centering
   \includegraphics[width=0.8\linewidth]{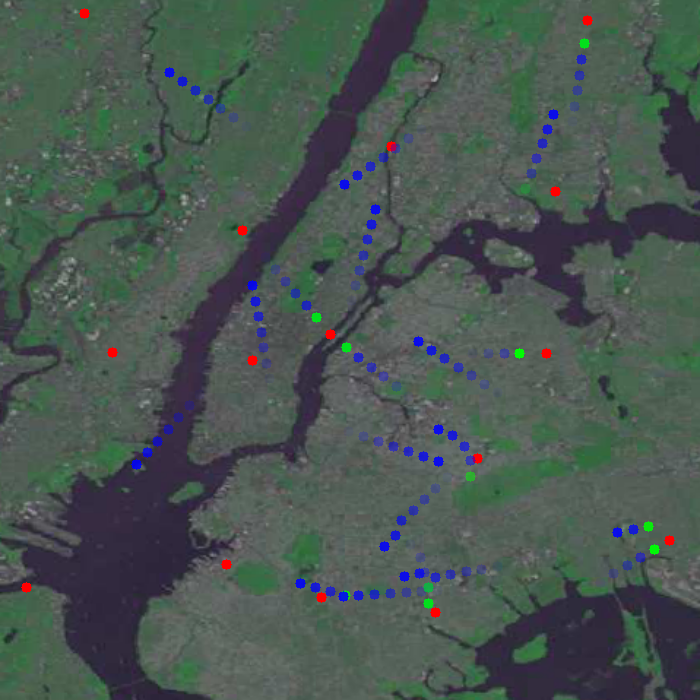}
   \caption{New York City Map}
   \label{fig:frame_nyc}
\end{subfigure}

\caption[Two numerical solutions]{Generated maps for the Bay Area and New York City. Red dots indicate vertiports and blue dots show the tracks of an agent's flight path over 60 seconds. Green dots indicate a landed agent. Note the area of the Bay Area map is larger than NYC.}
\label{fig:maps}
\end{figure}

The simulator generates hypothetical vertiport maps across a 120 km square of the San Francisco Bay Area and a 40 km square of New York City. Using United States population data from the 2020 Gridded Population of the World \cite{griddedpop}, we generate vertiport maps according to Algorithm \ref{alg:generate_vps}. Examples of maps are shown in Figure \ref{fig:maps}. Given a desired number of vertiports $m$ for the map, the average population served by each vertiport is calculated. Iteratively, a vertiport location is sampled according to the remaining population density distribution. The minimum circle centered at the sampled vertiport location required to contain the desired average population per vertiport is calculated, and the population contained within that circle is zeroed for the next iteration, having effectively been "served" by the newly generated vertiport. This process allows for a higher density of vertiports to be generated for urban areas, where the containing circles are relatively small, while suburban and rural areas have lower density vertiports. The purpose of this random generation is to provide a realistic test bed for the proposed ATNP solver while acknowledging that exact vertiport locations are not currently known. We argue that if the proposed method is effective over a range of randomly generated maps grounded in true population data then we should expect the system to perform well in future real instantiations of vertiport layouts.

% \usepackage{multirow}
\begin{table*}[]
\hskip-0.3cm
\begin{tabular}{lclrrrlrrr}
\toprule
                                         & \multicolumn{1}{l}{Assignment} &  & \multicolumn{7}{c}{Bay Area} \\ \cmidrule{1-2} \cmidrule{4-10} 
Trajectory Type                          &                                     &  & \multicolumn{3}{c}{MCTS}                                                 &  & \multicolumn{3}{c}{Greedy}                                                  \\ \cmidrule{1-2} \cmidrule{4-6} \cmidrule{8-10} 
Num agents                               &                                     &  & \multicolumn{1}{c}{10} & \multicolumn{1}{c}{40} & \multicolumn{1}{c}{80} &  & \multicolumn{1}{c}{100} & \multicolumn{1}{c}{200} & \multicolumn{1}{c}{300} \\
Num vertiports                           &                                     &  & \multicolumn{1}{c}{5}  & \multicolumn{1}{c}{12} & \multicolumn{1}{c}{16} &  & \multicolumn{1}{c}{20}  & \multicolumn{1}{c}{20}  & \multicolumn{1}{c}{20}  \\ \cmidrule{1-2} \cmidrule{4-6} \cmidrule{8-10} 
\multirow{3}{*}{NMACs / (hr agent)}      & Greedy                              &  & 0.000 $\pm$ 0.000       & 0.008 $\pm$ 0.011       & 0.025 $\pm$ 0.014       &  & 214.0 $\pm$ 185.7        & 232.3 $\pm$ 164.2        & 213.0 $\pm$ 41.5         \\
                                         & First-dispatch                      &  & 0.020 $\pm$ 0.040       & 0.008 $\pm$ 0.016       & 0.004 $\pm$ 0.009       &  & 1.858 $\pm$ 1.199        & 3.410 $\pm$ 0.608        & 3.479 $\pm$ 0.870        \\
                                         & Proposed                            &  & 0.040 $\pm$ 0.066       & 0.008 $\pm$ 0.011       & 0.003 $\pm$ 0.007       &  & 0.446 $\pm$ 0.079        & 0.989 $\pm$ 0.123        & 1.628 $\pm$ 0.28         \\ \cmidrule{1-2} \cmidrule{4-6} \cmidrule{8-10} 
\multirow{3}{*}{LOSs / (hr agent)}       & Greedy                              &  & 0.160 $\pm$ 0.102       & 0.538 $\pm$ 0.223       & 1.425 $\pm$ 0.275       &  & 340.8 $\pm$ 239.7        & 404.0 $\pm$ 220.5        & 425.9 $\pm$ 47.9         \\
                                         & First-dispatch                      &  & 0.220 $\pm$ 0.108       & 0.170 $\pm$ 0.071       & 0.152 $\pm$ 0.040       &  & 19.086 $\pm$ 1.927       & 35.649 $\pm$ 3.754       & 36.949 $\pm$ 10.309      \\
                                         & Proposed                            &  & 0.180 $\pm$ 0.160       & 0.092 $\pm$ 0.030       & 0.134 $\pm$ 0.032       &  & 13.492 $\pm$ 0.941       & 28.943 $\pm$ 1.840       & 30.423 $\pm$ 8.423       \\ \cmidrule{1-2} \cmidrule{4-6} \cmidrule{8-10} 
\multirow{3}{*}{Passengers / (hr agent)} & Greedy                              &  & 3.1 $\pm$ 0.5           & 3.1 $\pm$ 0.1           & 3.0 $\pm$ 0.4           &  & 3.0 $\pm$ 0.1            & 3.0 $\pm$ 0.1            & 2.9 $\pm$ 0.1            \\
                                         & First-dispatch                      &  & 3.2 $\pm$ 0.4           & 3.2 $\pm$ 0.2           & 3.0 $\pm$ 0.7           &  & 3.4 $\pm$ 0.1            & 3.1 $\pm$ 0.7            & 2.2 $\pm$ 1.0            \\
                                         & Proposed                            &  & 3.2 $\pm$ 0.3           & 2.7 $\pm$ 0.5           & 2.7 $\pm$ 0.5           &  & 2.8 $\pm$ 0.3            & 2.0 $\pm$ 0.2            & 1.8 $\pm$ 0.1            \\ \cmidrule{1-2} \cmidrule{4-6} \cmidrule{8-10} 
\multirow{3}{*}{Avg wait time}           & Greedy                              &  & 626 $\pm$ 344           & 479 $\pm$ 266           & 667 $\pm$ 41            &  & 865 $\pm$ 182            & 971 $\pm$ 290            & 921 $\pm$ 218            \\
                                         & First-dispatch                      &  & 708 $\pm$ 300           & 505 $\pm$ 468           & 804 $\pm$ 1174          &  & 279 $\pm$ 78             & 489 $\pm$ 541            & 1778 $\pm$ 1478          \\
                                     & Proposed                            &  & 4848 $\pm$ 292          & 203 $\pm$ 78            & 379 $\pm$ 306           &  & 245 $\pm$ 33             & 424 $\pm$ 195            & 511 $\pm$ 214            \\ \cmidrule{1-2} \cmidrule{4-6} \cmidrule{8-10} 
\multirow{3}{*}{Max wait time}           & Greedy                              &  & 2283 $\pm$ 990          & 2469 $\pm$ 1022         & 4400 $\pm$ 6234         &  & 5218 $\pm$ 838           & 4850 $\pm$ 843           & 5714 $\pm$ 679           \\
                                         & First-dispatch                      &  & 2355 $\pm$ 1029         & 2389 $\pm$ 1303         & 3962 $\pm$ 5525         &  & 1590 $\pm$ 433           & 3678 $\pm$ 4992          & 9276 $\pm$ 6775          \\
                                         & Proposed                            &  & 1625 $\pm$ 846          & 912 $\pm$ 298           & 2188 $\pm$ 1924         &  & 1258 $\pm$ 368           & 2710 $\pm$ 1523          & 3433 $\pm$ 1410          \\ \cmidrule{1-2} \cmidrule{4-6} \cmidrule{8-10} 
\multirow{3}{*}{Trip ratio}              & Greedy                              &  & 1.054 $\pm$ 0.062       & 1.089 $\pm$ 0.130       & 1.122 $\pm$ 0.208       &  & 1.064 $\pm$ 0.104        & 1.069 $\pm$ 0.122        & 1.068 $\pm$ 0.119        \\
                                         & First-dispatch                      &  & 1.052 $\pm$ 0.062       & 1.071 $\pm$ 0.100       & 1.087 $\pm$ 0.126       &  & 1.066 $\pm$ 0.108        & 1.069 $\pm$ 0.122        & 1.064 $\pm$ 0.102        \\
                                         & Proposed                            &  & 1.074 $\pm$ 0.204       & 1.139 $\pm$ 0.301       & 1.182 $\pm$ 0.373       &  & 1.183 $\pm$ 0.449        & 1.276 $\pm$ 1.018        & 1.367 $\pm$ 1.596        \\ \bottomrule
\end{tabular}
\caption{Simulation results for the Bay Area map.}
\label{fig:results_sf}
\end{table*}

% \usepackage{multirow}
\begin{table*}[]
\centering
\begin{tabular}{lclrrrlrr}
\toprule
                                         & \multicolumn{1}{l}{Assignment} &  & \multicolumn{6}{c}{New York City}                                                                                               \\ \cmidrule{1-2} \cmidrule{4-9} 
Trajectory Type                          &                                     &  & \multicolumn{3}{c}{MCTS}                                                 &  & \multicolumn{2}{c}{Greedy}                        \\ \cmidrule{1-2} \cmidrule{4-6} \cmidrule{8-9} 
Num agents                               &                                     &  & \multicolumn{1}{c}{10} & \multicolumn{1}{c}{20} & \multicolumn{1}{c}{40} &  & \multicolumn{1}{c}{100} & \multicolumn{1}{c}{200} \\
Num vertiports                           &                                     &  & \multicolumn{1}{c}{5}  & \multicolumn{1}{c}{8}  & \multicolumn{1}{c}{12} &  & \multicolumn{1}{c}{16}  & \multicolumn{1}{c}{16}  \\ \cmidrule{1-2} \cmidrule{4-6} \cmidrule{8-9} 
\multirow{3}{*}{NMACs / (hr agent)}      & Greedy                              &  & 0.000 $\pm$ 0.000       & 0.005 $\pm$ 0.015       & 0.005 $\pm$ 0.010       &  & 12.73 $\pm$ 1.971        & 24.16 $\pm$ 3.910        \\
                                         & First-dispatch                      &  & 0.000 $\pm$ 0.000       & 0.005 $\pm$ 0.015       & 0.012 $\pm$ 0.017       &  & 1.055 $\pm$ 0.354        & 1.906 $\pm$ 0.190        \\
                                         & Proposed                            &  & 0.000 $\pm$ 0.000       & 0.005 $\pm$ 0.015       & 0.012 $\pm$ 0.023       &  & 0.506 $\pm$ 0.089        & 1.096 $\pm$ 0.091        \\ \cmidrule{1-2} \cmidrule{4-6} \cmidrule{8-9} 
\multirow{3}{*}{LOSs / (hr agent)}       & Greedy                              &  & 0.110 $\pm$ 0.070       & 0.120 $\pm$ 0.117       & 0.370 $\pm$ 0.175       &  & 49.57 $\pm$ 5.927        & 90.71 $\pm$ 6.944        \\
                                         & First-dispatch                      &  & 0.090 $\pm$ 0.070       & 0.090 $\pm$ 0.089       & 0.220 $\pm$ 0.094       &  & 23.59 $\pm$ 1.03         & 42.07 $\pm$ 1.90         \\
                                         & Proposed                            &  & 0.040 $\pm$ 0.049       & 0.065 $\pm$ 0.032       & 0.182 $\pm$ 0.053       &  & 20.00 $\pm$ 0.908        & 37.97 $\pm$ 1.041        \\ \cmidrule{1-2} \cmidrule{4-6} \cmidrule{8-9} 
\multirow{3}{*}{Passengers / (hr agent)} & Greedy                              &  & 9.6 $\pm$ 1.1           & 9.6 $\pm$ 0.8           & 9.6 $\pm$ 0.4           &  & 10.3 $\pm$ 0.5           & 10.1 $\pm$ 0.3           \\
                                         & First-dispatch                      &  & 9.8 $\pm$ 0.8           & 9.2 $\pm$ 0.5           & 9.5 $\pm$ 0.4           &  & 10.5 $\pm$ 0.4           & 10.1 $\pm$ 0.3           \\
                                         & Proposed                            &  & 9.2 $\pm$ 2.6           & 9.9 $\pm$ 0.7           & 8.7 $\pm$ 2.3           &  & 7.9 $\pm$ 2.2            & 7.8 $\pm$ 0.3            \\ \cmidrule{1-2} \cmidrule{4-6} \cmidrule{8-9} 
\multirow{3}{*}{Avg wait time (s)}       & Greedy                              &  & 302 $\pm$ 120           & 259 $\pm$ 97            & 188 $\pm$ 43            &  & 106 $\pm$ 22             & 77 $\pm$ 36              \\
                                         & First-dispatch                      &  & 340 $\pm$ 106           & 352 $\pm$ 122           & 285 $\pm$ 70            &  & 111 $\pm$ 33             & 66 $\pm$ 9               \\
                                         & Proposed                            &  & 251 $\pm$ 88            & 195 $\pm$ 91            & 165 $\pm$ 51            &  & 107 $\pm$ 15             & 242 $\pm$ 18             \\ \cmidrule{1-2} \cmidrule{4-6} \cmidrule{8-9} 
\multirow{3}{*}{Max wait time (s)}       & Greedy                              &  & 1004 $\pm$ 302          & 1025 $\pm$ 430          & 875 $\pm$ 251           &  & 616 $\pm$ 139            & 474 $\pm$ 181            \\
                                         & First-dispatch                      &  & 955 $\pm$ 251           & 1274 $\pm$ 404          & 1167 $\pm$ 248          &  & 561 $\pm$ 184            & 308 $\pm$ 68             \\
                                         & Proposed                            &  & 781 $\pm$ 173           & 638 $\pm$ 332           & 643 $\pm$ 300           &  & 404 $\pm$ 74             & 936 $\pm$ 84             \\ \cmidrule{1-2} \cmidrule{4-6} \cmidrule{8-9} 
\multirow{3}{*}{Trip ratio}              & Greedy                              &  & 1.170 $\pm$ 0.160       & 1.259 $\pm$ 0.450       & 1.369 $\pm$ 0.543       &  & 1.205 $\pm$ 0.271        & 1.204 $\pm$ 0.273        \\
                                         & First-dispatch                      &  & 1.164 $\pm$ 0.184       & 1.226 $\pm$ 0.314       & 1.355 $\pm$ 0.637       &  & 1.204 $\pm$ 0.251        & 1.203 $\pm$ 0.269        \\
                                         & Proposed                            &  & 1.156 $\pm$ 0.161       & 1.226 $\pm$ 0.265       & 1.391 $\pm$ 0.542       &  & 1.495 $\pm$ 1.273        & 1.898 $\pm$ 2.422        \\ \bottomrule
\end{tabular}
\caption{Simulation results for the New York City map.}
\label{fig:results_nyc}
\end{table*}

\begin{table}[]
\begin{tabular}{lccl}
\toprule
                   & Flight levels & Selection type &                  \\ \midrule
NMACs / (hr agent) & 2                 & Random         & 0.391 $\pm$ 0.076 \\
                   &                   & Proposed       & 0.209 $\pm$ 0.040 \\ \cmidrule{2-4} 
                   & 3                 & Random         & 0.259 $\pm$ 0.070 \\
                   &                   & Proposed       & 0.132 $\pm$ 0.024 \\ \cmidrule{2-4} 
                   & 4                 & Random         & 0.190 $\pm$ 0.058 \\
                   &                   & Proposed       & 0.095 $\pm$ 0.045 \\ \midrule
LOSs / (hr agent)  & 2                 & Random         & 9.609 $\pm$ 0.600 \\
                   &                   & Proposed       & 8.362 $\pm$ 0.454 \\ \cmidrule{2-4} 
                   & 3                 & Random         & 6.417 $\pm$ 0.555 \\
                   &                   & Proposed       & 5.053 $\pm$ 0.255 \\ \cmidrule{2-4} 
                   & 4                 & Random         & 4.824 $\pm$ 0.357 \\
                   &                   & Proposed       & 3.467 $\pm$ 0.206 \\ \bottomrule
\end{tabular}
\caption{NMAC and LOS rates for various numbers of flight levels in a map of 100 agents and 16 vertiports across NYC. In random selection, a flight level is selected at random for each agent. In the proposed method, a the system considers future airspace density before assigning a level or holding the aircraft for traffic to pass.}\label{fig:flight_levels}
\end{table}

\subsection{Evaluation Metrics}
The proposed solution to the ATNP is evaluated in terms of safety, operational efficiency, and trip efficiency, with specific metrics described in the following subsections.

\subsubsection{Safety}
Safety is evaluated in terms of near mid-air collisions, which is when two aircraft are within 500 feet (0.15 km) of each other, and loss of separation events, which is when two aircraft are within 0.5 nautical miles (0.926 km). The occurrence of these events is normalized by hour and number of agents.

\subsubsection{Operational Efficiency}
Passenger waiting times at vertiports are important because long wait times negatively affect the customer experience. This is assessed as the average and maximum passenger wait times for a given simulation. Likewise, the number of passengers delivered per hour per agent provides a throughput efficiency metric which characterizes how the proposed methods perform across the varying number of agents used in the experiments.

\subsubsection{Trip Efficiency}
Intuitively, it is preferred that each UAV flies as close to the optimal route between source and destination vertiports as possible, while maintaining a safe flight path, in order to reduce passenger travel times and increase throughput of the overall UAV network. This intuition suggests an efficiency metric, which can be calculated as the realized flight time divided by the minimum possible flight time for a given route. While trip efficiency is not directly considered in the reward model, it is expected to emerge as a consequence of the objective that maximizes passenger throughput.

\subsection{Experiment Parameters}
We consider a varying number of $n$ eVTOLs and $m$ vertiports in our experiments. In each simulation, we generate $10n$ passengers and the simulation is complete when all the passengers are delivered. Each experimental condition is run 10 times with a different random seed, which changes the locations of the generated vertiports. We determine a nominal network-wide arrival rate for the passengers by setting an average trip time as the time it takes an agent to travel $\frac{2}{3}$ the map distance, then calculating the average trips per hour per agent, and finally scaling that by the number of agents. Individual vertiport arrival rates are assigned from the network-wide arrival rate in proportion to a vertiport's population density.

We consider timesteps of $\Delta t = 10$ seconds, a constant eVTOL velocity of 90 m/s, and a vertiport landing radius of $d_\text{land} = 1.7$ km. The MCTS action space consists of three angular rates $\omega \in \{ -0.04, 0, 0.04\}$ flying and four takeoff angles $\theta_\text{takeoff} \in \{0, \frac{\pi}{2}, \pi, \frac{3\pi}{2} \}$ when grounded. For the candidate matching generation we consider the top $k = 10$ matches. The lookahead time for flight level selection is 200 seconds and the lookahead for determining if agents require MCTS is $\Delta t_{\text{lookahead}} = 60$ seconds. MCTS is run for 50 iterations with a maximum depth of 4.

The proposed method is compared against two baselines: a simple greedy agent-passenger assignment method that at each timestep simply assigns an agent to the nearest passenger, and the first-dispatch protocol, which uses the Hungarian algorithm to assign agents to passengers. However, in contrast to the proposed method which dynamically reassigns agents as new passengers enter the system, the first-dispatch protocol does not change assignments once given. 

\begin{algorithm}
\caption{\textsc{GenerateVertiports}}\label{alg:generate_vps}
\textbf{Input: } number of vertiports $m$, 2D population density grid P\\
\textbf{Output: } $m$ vertiport locations
\begin{algorithmic}[1]

\STATE $\texttt{locations} \gets \{ \}$
\STATE $\texttt{vp\_pop} \gets \textsc{Sum}(P) / m$
\FOR{$k \in \{1, 2, \dots, m\}$}
    \STATE $\texttt{dist} \gets P / \textsc{Sum}(P)$
    \STATE $(x, y) \gets \textsc{Sample}(\texttt{dist})$
    \STATE $r \gets \textsc{MinimumCircle}(x, y, P)$
    \STATE $P_{\textsc{Circle}(x, y, r)} \gets 0$
\ENDFOR

\RETURN $\texttt{locations}$
\end{algorithmic}
\end{algorithm}

\section{Results}
Figure \ref{fig:results_sf} shows the primary results for the Bay Area map and Figure \ref{fig:results_nyc} shows the primary results for the New York City map. Each of these sets of experiments only allowed for a single flight level. Due to the low computation speed of the Python simulator, greedy trajectory planning was used for $n \ge 100$ agents. The results show that MCTS trajectory planning drastically reduces the rate of NMAC and LOS event. The proposed method reduces both the average and maximum passenger wait times compared to the baselines, which is a result of dynamic reassignment. Furthermore, the proposed method reduces the rate of NMAC and LOS events, which is a result of two empirical phenomena observed in the simulations. Firstly, the greedy assignment method results in many redundant aircraft heading towards the same destination, resulting in high denisty air traffic. The proposed method attempts to disperse traffic across the map to where it is required through the distribution matching selection process. Secondly, the flight level selection phase holds an aircraft at the vertiport if there is high aircraft density along its assigned route. This prevents many potential collisions before flight trajectory planning is even invoked. This holding does slightly decrease the passenger throughput of the system and results in slightly higher trip ratios when compared to the immediate takeoff baselines.

Figure \ref{fig:flight_levels} shows how the proposed flight selection method performs in terms of safety against a random flight level policy in a map of 100 agents and 16 vertiports across NYC. As the number of flight levels increases, the NMAC and LOS event rates decrease as expected, with the proposed method having uniformly lower rates compared to the random policy. 

\section{Conclusion}
This work introduces the air taxi network problem, which models the optimization problem of future cooperative air taxi networks. A solution to this problem would help enable autonomous, high density, and dynamic air traffic control systems. We propose a three-phase decision making model to solve the ATNP and perform simulations that show large improvements in safety and passenger wait times over greedy and first-dispatch method baselines. Future work can consider more complex vertiport models that include finite landing pads, passenger loading times, and time-varying passenger demand. Similarly, future work can allow agents to have full 3D free flight and a battery model to better align with real-world constraints.

\bibliographystyle{IEEEtran}
\bibliography{bib}

\begin{thebibliography}{10}
\providecommand{\url}[1]{#1}
\csname url@samestyle\endcsname
\providecommand{\newblock}{\relax}
\providecommand{\bibinfo}[2]{#2}
\providecommand{\BIBentrySTDinterwordspacing}{\spaceskip=0pt\relax}
\providecommand{\BIBentryALTinterwordstretchfactor}{4}
\providecommand{\BIBentryALTinterwordspacing}{\spaceskip=\fontdimen2\font plus
\BIBentryALTinterwordstretchfactor\fontdimen3\font minus \fontdimen4\font\relax}
\providecommand{\BIBforeignlanguage}[2]{{%
\expandafter\ifx\csname l@#1\endcsname\relax
\typeout{** WARNING: IEEEtran.bst: No hyphenation pattern has been}%
\typeout{** loaded for the language `#1'. Using the pattern for}%
\typeout{** the default language instead.}%
\else
\language=\csname l@#1\endcsname
\fi
#2}}
\providecommand{\BIBdecl}{\relax}
\BIBdecl

\bibitem{owid-urbanization}
H.~Ritchie and M.~Roser, ``Urbanization,'' \emph{Our World in Data}, 2019, https://ourworldindata.org/urbanization.

\bibitem{Inrix_2023}
\BIBentryALTinterwordspacing
Inrix, ``Return to work, higher gas prices \& inflation drove americans to spend hundreds more in time and money commuting,'' Nov 2023. [Online]. Available: \url{https://inrix.com/press-releases/2022-global-traffic-scorecard-us/}
\BIBentrySTDinterwordspacing

\bibitem{mckinsey}
\BIBentryALTinterwordspacing
\emph{The future of air mobility: Electric aircraft and Flying Taxis}.\hskip 1em plus 0.5em minus 0.4em\relax McKinsey \& Company, Nov. 2021. [Online]. Available: \url{https://www.mckinsey.com}
\BIBentrySTDinterwordspacing

\bibitem{Goodrich2021}
\BIBentryALTinterwordspacing
K.~H. Goodrich and C.~R. Theodore, ``Description of the nasa urban air mobility maturity level (uml) scale,'' in \emph{AIAA Scitech 2021 Forum}.\hskip 1em plus 0.5em minus 0.4em\relax American Institute of Aeronautics and Astronautics, Jan. 2021. [Online]. Available: \url{http://dx.doi.org/10.2514/6.2021-1627}
\BIBentrySTDinterwordspacing

\bibitem{Xiang2023}
\BIBentryALTinterwordspacing
S.~Xiang, A.~Xie, M.~Ye, X.~Yan, X.~Han, H.~Niu, Q.~Li, and H.~Huang, ``Autonomous evtol: A summary of researches and challenges,'' \emph{Green Energy and Intelligent Transportation}, p. 100140, Nov. 2023. [Online]. Available: \url{http://dx.doi.org/10.1016/j.geits.2023.100140}
\BIBentrySTDinterwordspacing

\bibitem{nextgen}
\BIBentryALTinterwordspacing
\emph{NextGen Implementation Plan}, 2018. [Online]. Available: \url{https://www.faa.gov}
\BIBentrySTDinterwordspacing

\bibitem{nextgen_evtol}
\BIBentryALTinterwordspacing
\emph{Urban Air Mobility (UAM) Concept of Operations}.\hskip 1em plus 0.5em minus 0.4em\relax Federal Aviation Administration, Apr 2023. [Online]. Available: \url{https://www.faa.gov/air-taxis/uam\_blueprint}
\BIBentrySTDinterwordspacing

\bibitem{Vascik2020SystemsAO}
\BIBentryALTinterwordspacing
P.~D. Vascik, ``Systems analysis of urban air mobility operational scaling,'' 2020. [Online]. Available: \url{https://api.semanticscholar.org/CorpusID:213325960}
\BIBentrySTDinterwordspacing

\bibitem{Rajendran2020}
\BIBentryALTinterwordspacing
S.~Rajendran and S.~Srinivas, ``Air taxi service for urban mobility: A critical review of recent developments, future challenges, and opportunities,'' \emph{Transportation Research Part E: Logistics and Transportation Review}, vol. 143, p. 102090, Nov. 2020. [Online]. Available: \url{http://dx.doi.org/10.1016/j.tre.2020.102090}
\BIBentrySTDinterwordspacing

\bibitem{ubermatching}
\BIBentryALTinterwordspacing
C.~Yan, H.~Zhu, N.~Korolko, and D.~Woodard, ``Dynamic pricing and matching in ride‐hailing platforms,'' \emph{Naval Research Logistics (NRL)}, vol.~67, no.~8, p. 705–724, Nov. 2019. [Online]. Available: \url{http://dx.doi.org/10.1002/nav.21872}
\BIBentrySTDinterwordspacing

\bibitem{vrp}
\BIBentryALTinterwordspacing
\emph{The Vehicle Routing Problem}.\hskip 1em plus 0.5em minus 0.4em\relax Society for Industrial and Applied Mathematics, Jan. 2002. [Online]. Available: \url{http://dx.doi.org/10.1137/1.9780898718515}
\BIBentrySTDinterwordspacing

\bibitem{Wang2019}
\BIBentryALTinterwordspacing
Z.~Wang and J.-B. Sheu, ``Vehicle routing problem with drones,'' \emph{Transportation Research Part B: Methodological}, vol. 122, p. 350–364, Apr. 2019. [Online]. Available: \url{http://dx.doi.org/10.1016/j.trb.2019.03.005}
\BIBentrySTDinterwordspacing

\bibitem{9197313}
S.~Choudhury, K.~Solovey, M.~J. Kochenderfer, and M.~Pavone, ``Efficient large-scale multi-drone delivery using transit networks,'' in \emph{2020 IEEE International Conference on Robotics and Automation (ICRA)}, 2020, pp. 4543--4550.

\bibitem{Han2023}
\BIBentryALTinterwordspacing
J.~Han, Y.~Liu, and Y.~Li, ``Vehicle routing problem with drones considering time windows and dynamic demand,'' \emph{Applied Sciences}, vol.~13, no.~24, p. 13086, Dec. 2023. [Online]. Available: \url{http://dx.doi.org/10.3390/app132413086}
\BIBentrySTDinterwordspacing

\bibitem{Yang2020ScalableMC}
\BIBentryALTinterwordspacing
X.~Yang and P.~Wei, ``Scalable multi-agent computational guidance with separation assurance for autonomous urban air mobility,'' \emph{Journal of Guidance Control and Dynamics}, vol.~43, pp. 1473--1486, 2020. [Online]. Available: \url{https://api.semanticscholar.org/CorpusID:219435229}
\BIBentrySTDinterwordspacing

\bibitem{Kucharska2019}
\BIBentryALTinterwordspacing
Kucharska, ``Dynamic vehicle routing problem—predictive and unexpected customer availability,'' \emph{Symmetry}, vol.~11, no.~4, p. 546, Apr. 2019. [Online]. Available: \url{http://dx.doi.org/10.3390/sym11040546}
\BIBentrySTDinterwordspacing

\bibitem{OjedaRios2021}
\BIBentryALTinterwordspacing
B.~H. Ojeda~Rios, E.~C. Xavier, F.~K. Miyazawa, P.~Amorim, E.~Curcio, and M.~J. Santos, ``Recent dynamic vehicle routing problems: A survey,'' \emph{Computers; Industrial Engineering}, vol. 160, p. 107604, Oct. 2021. [Online]. Available: \url{http://dx.doi.org/10.1016/j.cie.2021.107604}
\BIBentrySTDinterwordspacing

\bibitem{mapf-kc}
W.~Hoenig, T.~Kumar, L.~Cohen, H.~Ma, H.~Xu, N.~Ayanian, and S.~Koenig, ``Multi-agent path finding with kinematic constraints,'' 06 2016.

\bibitem{Bai2011}
\BIBentryALTinterwordspacing
H.~Bai, D.~Hsu, M.~Kochenderfer, and W.~Sun~Lee, ``Unmanned aircraft collision avoidance using continuous-state pomdps,'' in \emph{Robotics: Science and Systems VII}, ser. RSS2011.\hskip 1em plus 0.5em minus 0.4em\relax Robotics: Science and Systems Foundation, Jun. 2011. [Online]. Available: \url{http://dx.doi.org/10.15607/RSS.2011.VII.001}
\BIBentrySTDinterwordspacing

\bibitem{Wolf2011}
\BIBentryALTinterwordspacing
T.~B. Wolf and M.~J. Kochenderfer, ``Aircraft collision avoidance using monte carlo real-time belief space search,'' \emph{Journal of Intelligent Robotic Systems}, vol.~64, no.~2, p. 277–298, Jan. 2011. [Online]. Available: \url{http://dx.doi.org/10.1007/s10846-010-9532-6}
\BIBentrySTDinterwordspacing

\bibitem{Mueller2016}
\BIBentryALTinterwordspacing
E.~R. Mueller and M.~Kochenderfer, ``Multi-rotor aircraft collision avoidance using partially observable markov decision processes,'' in \emph{AIAA Modeling and Simulation Technologies Conference}.\hskip 1em plus 0.5em minus 0.4em\relax American Institute of Aeronautics and Astronautics, Jun. 2016. [Online]. Available: \url{http://dx.doi.org/10.2514/6.2016-3673}
\BIBentrySTDinterwordspacing

\bibitem{acasx2}
M.~Kochenderfer, J.~Holland, and J.~Chryssanthacopoulos, ``Next generation airborne collision avoidance system,'' \emph{Lincoln Laboratory Journal}, vol.~19, pp. 17--33, 01 2012.

\bibitem{acasx}
\BIBentryALTinterwordspacing
\emph{ACAS X}.\hskip 1em plus 0.5em minus 0.4em\relax SKYbrary Aviation Safety. [Online]. Available: \url{https://skybrary.aero/articles/acas-x}
\BIBentrySTDinterwordspacing

\bibitem{Ong2017}
\BIBentryALTinterwordspacing
H.~Y. Ong and M.~J. Kochenderfer, ``Markov decision process-based distributed conflict resolution for drone air traffic management,'' \emph{Journal of Guidance, Control, and Dynamics}, vol.~40, no.~1, p. 69–80, Jan. 2017. [Online]. Available: \url{http://dx.doi.org/10.2514/1.G001822}
\BIBentrySTDinterwordspacing

\bibitem{Zhao2020OnlineMC}
\BIBentryALTinterwordspacing
P.~Zhao, W.~Wang, L.~Ying, B.~Sridhar, and Y.~Liu, ``Online multiple-aircraft collision avoidance method,'' \emph{Journal of Guidance Control and Dynamics}, vol.~43, pp. 1456--1472, 2020. [Online]. Available: \url{https://api.semanticscholar.org/CorpusID:225680205}
\BIBentrySTDinterwordspacing

\bibitem{huang2022multiuav}
S.~Huang, H.~Zhang, and Z.~Huang, ``Multi-uav collision avoidance using multi-agent reinforcement learning with counterfactual credit assignment,'' 2022.

\bibitem{Chen2016DecentralizedNM}
\BIBentryALTinterwordspacing
Y.~F. Chen, M.~Liu, M.~Everett, and J.~P. How, ``Decentralized non-communicating multiagent collision avoidance with deep reinforcement learning,'' \emph{2017 IEEE International Conference on Robotics and Automation (ICRA)}, pp. 285--292, 2016. [Online]. Available: \url{https://api.semanticscholar.org/CorpusID:8342451}
\BIBentrySTDinterwordspacing

\bibitem{equipadsb}
\BIBentryALTinterwordspacing
\emph{Equip ADS-B}.\hskip 1em plus 0.5em minus 0.4em\relax Federal Aviation Administration. [Online]. Available: \url{https://www.faa.gov/air\_traffic/technology/equipadsb}
\BIBentrySTDinterwordspacing

\bibitem{adsbspec}
\BIBentryALTinterwordspacing
``A component-level model of automatic dependent surveillance - broadcast (ads-b),'' Jun 2018. [Online]. Available: \url{https://ntrs.nasa.gov}
\BIBentrySTDinterwordspacing

\bibitem{Zaid2023}
\BIBentryALTinterwordspacing
A.~A. Zaid, B.~E.~Y. Belmekki, and M.-S. Alouini, ``evtol communications and networking in uam: Requirements, key enablers, and challenges,'' \emph{IEEE Communications Magazine}, vol.~61, no.~8, p. 154–160, Aug. 2023. [Online]. Available: \url{http://dx.doi.org/10.1109/MCOM.004.2300061}
\BIBentrySTDinterwordspacing

\bibitem{Zhang2017}
\BIBentryALTinterwordspacing
L.~Zhang, T.~Hu, Y.~Min, G.~Wu, J.~Zhang, P.~Feng, P.~Gong, and J.~Ye, ``A taxi order dispatch model based on combinatorial optimization,'' in \emph{Proceedings of the 23rd ACM SIGKDD International Conference on Knowledge Discovery and Data Mining}, ser. KDD ’17.\hskip 1em plus 0.5em minus 0.4em\relax ACM, Aug. 2017. [Online]. Available: \url{http://dx.doi.org/10.1145/3097983.3098138}
\BIBentrySTDinterwordspacing

\bibitem{ashlagi2018maximum}
I.~Ashlagi, M.~Burq, C.~Dutta, P.~Jaillet, A.~Saberi, and C.~Sholley, ``Maximum weight online matching with deadlines,'' 2018.

\bibitem{Kuhn1955}
\BIBentryALTinterwordspacing
H.~W. Kuhn, ``The hungarian method for the assignment problem,'' \emph{Naval Research Logistics Quarterly}, vol.~2, no. 1–2, p. 83–97, Mar. 1955. [Online]. Available: \url{http://dx.doi.org/10.1002/nav.3800020109}
\BIBentrySTDinterwordspacing

\bibitem{zkan2020}
\BIBentryALTinterwordspacing
E.~\"{O}zkan and A.~R. Ward, ``Dynamic matching for real-time ride sharing,'' \emph{Stochastic Systems}, vol.~10, no.~1, p. 29–70, Mar. 2020. [Online]. Available: \url{http://dx.doi.org/10.1287/stsy.2019.0037}
\BIBentrySTDinterwordspacing

\bibitem{8569645}
I.~C. Kleinbekman, M.~A. Mitici, and P.~Wei, ``evtol arrival sequencing and scheduling for on-demand urban air mobility,'' in \emph{2018 IEEE/AIAA 37th Digital Avionics Systems Conference (DASC)}, 2018, pp. 1--7.

\bibitem{Song2022}
\BIBentryALTinterwordspacing
K.~Song, ``Optimal vertiport airspace and approach control strategy for urban air mobility (uam),'' \emph{Sustainability}, vol.~15, no.~1, p. 437, Dec. 2022. [Online]. Available: \url{http://dx.doi.org/10.3390/su15010437}
\BIBentrySTDinterwordspacing

\bibitem{Li2020}
\BIBentryALTinterwordspacing
S.~Li, M.~Egorov, and M.~J. Kochenderfer, ``Analysis of fleet management and infrastructure constraints in on-demand urban air mobility operations,'' in \emph{AIAA AVIATION 2020 FORUM}.\hskip 1em plus 0.5em minus 0.4em\relax American Institute of Aeronautics and Astronautics, Jun. 2020. [Online]. Available: \url{http://dx.doi.org/10.2514/6.2020-2907}
\BIBentrySTDinterwordspacing

\bibitem{mmdp}
C.~Boutilier, ``Planning, learning and coordination in multiagent decision processes,'' \emph{Proceedings of the 6th Conference on Theoretical Aspects of Rationality and Knowledge}, 02 1970.

\bibitem{532080bb-34c4-3c21-a5ce-6f11e19925ad}
\BIBentryALTinterwordspacing
K.~G. Murty, ``An algorithm for ranking all the assignments in order of increasing cost,'' \emph{Operations Research}, vol.~16, no.~3, pp. 682--687, 1968. [Online]. Available: \url{http://www.jstor.org/stable/168595}
\BIBentrySTDinterwordspacing

\bibitem{uct}
L.~Kocsis and C.~Szepesvári, ``Bandit based monte-carlo planning,'' vol. 2006, 09 2006, pp. 282--293.

\bibitem{griddedpop}
\BIBentryALTinterwordspacing
{Center For International Earth Science Information Network-CIESIN-Columbia University}, ``Gridded population of the world, version 4 (gpwv4): Population count, revision 11,'' 2018. [Online]. Available: \url{https://sedac.ciesin.columbia.edu/data/set/gpw-v4-population-count-rev11}
\BIBentrySTDinterwordspacing

\end{thebibliography}

\vspace{12pt}

\end{document}